\documentclass[conference]{IEEEtran}

\usepackage{cite}
\usepackage{pslatex} 
\usepackage[colorlinks=false,bookmarks=false]{hyperref}
\usepackage{booktabs}
\usepackage{graphicx}
\usepackage{xcolor}
\usepackage{multirow}
\usepackage{comment}
\usepackage{listings}
\usepackage{xspace}		
\usepackage{cleveref}	

\newcommand{\SV}{SystemVerilog\xspace}

\usepackage{listings}
\newcommand{\longlist}[3]{{\lstinputlisting[float, caption={#2}, label={#3}, frame=tb, captionpos=b]{#1}}}

\title{Open-Source Verification with Chisel and Scala}

\author{\IEEEauthorblockN{Andrew Dobis, Tjark Petersen, Kasper Juul Hesse Rasmussen, Enrico Tolotto, \\
Hans Jakob Damsgaard, Simon Thye Andersen, Richard Lin, Martin Schoeberl}\\
\IEEEauthorblockA{\textit{Department of Applied Mathematics and Computer Science} \\
\textit{Technical University of Denmark}\\
Lyngby, Denmark \\\\
\textit{Department of Electrical Engineering and Computer Sciences} \\
\textit{UC Berkeley}\\
Berkeley, CA \\\\
andrew.dobis@alumni.epfl.ch, s186083@student.dtu.dk, s183735@student.dtu.dk, s190057@student.dtu.dk, \\
s163915@student.dtu.dk, simon.thye@gmail.com, richard.lin@berkeley.edu, masca@dtu.dk}
}

\begin{document}

\maketitle \thispagestyle{empty}

\begin{abstract}
Performance increase with general-purpose processors has come to a halt.
We can no longer depend on Moore's Law to increase computing performance.
The only way to achieve higher performance or lower energy consumption
is by building domain-specific hardware accelerators.
To efficiently design and verify those domain-specific accelerators, we need
agile hardware development. One of the main obstacles when proposing such a modern method
is the lack of modern tools to attack it. To be able to verify a design in such a time-constrained development
method, one needs to have efficient tools both for design and verification.

This paper thus proposes ChiselVerify, an open-source tool for verifying
circuits described in any Hardware Description Language. It builds on top of the Chisel
hardware construction language and uses Scala to drive the verification using a testing strategy 
inspired by the Universal Verification Methodology (UVM) and adapted for designs described in Chisel.
ChiselVerify is created based on three key ideas.
First, our solution highly increases the productivity of the verification engineer, by allowing hardware testing to be done in a modern high-level programming environment.
Second, the framework functions with any hardware description language thanks to the flexibility of Chisel blackboxes.
Finally, the solution is well integrated into the existing Chisel universe, making it an extension of currently existing testing libraries.

We implement ChiselVerify in a way inspired by the functionalities found in SystemVerilog. This allows one to use
functional coverage, constrained-random verification, bus functional models, transaction-level modeling and much more
during the verification process of a design in a contemporary high-level programming ecosystem.
\end{abstract}

\begin{IEEEkeywords}
digital design, verification, Chisel, Scala
\end{IEEEkeywords}

\section{Introduction}
\label{sec:introduction}

We can no longer depend on Moore's Law to increase computing performance~\cite{dark-silicon:2011}.
Performance increase with general-purpose processors has come to a halt.
The only way to achieve higher performance or lower energy consumption
is by building domain-specific hardware accelerators~\cite{domain-hw-acc:2020}.
These accelerators can be built in chips or in FPGAs in the cloud.
The production of a chip is costly. Therefore, it is essential to get
the design right at the first tape-out. Thorough testing and verification of the design is mandatory.

To efficiently develop and verify those accelerators, we can learn from software development trends such as agile software development~\cite{agile:manifesto}.
We believe that we need to adapt to agile hardware development~\cite{henn-patt:turing:2019}.

Furthermore, as accelerators become part of the cloud service, i.e. FPGAs in the cloud,
software developers will increasingly need to adapt critical algorithms to FPGAs to enhance performance.
Hence, it is imperative to make accelerator design accessible for software developers.
By adapting hardware accelerator design to the methods and tools of contemporary software design,
it is possible to bridge both domains catering for a more uniform hardware/software development process.

Up until a few years ago, the two main design languages, Verilog and VHDL, dominated the
design and testing of digital circuits.
However, compared to software development and testing, digital design and testing methods/tools 
lack several decades of development. We propose a tool that
leverages software development and testing methods for digital design.
Based in the hardware construction language Chisel~\cite{chisel:dac2012}, which itself is embedded in Scala,
our framework reimagines functionalities form Universal Verification Method (UVM) with SystemVerilog~\cite{SystemVerilog} and
adapts them for the existing Chisel ecosystem.

We thus developed a method and concrete tools for agile hardware development.
ChiselVerify combines tools, languages, development, and testing methods from the last decades in
software development and applies them to hardware design.
We aim to raise the tooling level for a digital design to increase productivity.
The workload demanded for the verification (testing) of digital systems is about double the time of developing
them in the first place.

Using the power and flexibility of Chisel Blackboxes, our tool can be used to verify
designs implemented in any of the major hardware description languages (i.e., VHDL or Verilog)
with little overhead. Furthermore, golden models described in any programming language can be
used using the Java JNI. Our work builds upon existing open-source projects and 
is thus also open-source.

We developed an object-oriented and functional framework for verification in Scala.
This framework is inspired by UVM, but leverages Scala's conciseness with the
combination of object-oriented programming with functional programming.
An initial experiment of testing the accumulator circuit of the Leros processor~\cite{leros:arcs2019}
showed the that a test written with UVM was about 800 lines of code, where a Scala based
test was around 80 lines of code~\cite{verify:chisel:2020}.
However, UVM supports more functionality that a plain ChiselTest in Scala.

Within our verification framework, we support mixed language verification.
Verilog can easily be combined with Chisel, as Chisel generates Verilog, and
we use ChiselTest as a driver for the open-source Verilog simulator Verlator.
With Yosys synthesis suite~\cite{Yosys} and GHDL~\cite{ghdl}
we can translate VHDL into Verilog.

A verification method is only usable when it can handle mixed-source designs.
This means a Scala driven method must be able to test components written in Verilog,
VHDL, and SystemVerilog.

Chisel has support for black boxes, which allows the use of Verilog code within the Chisel design.
Therefore, it is relatively easy to integrate Verilog components when wrapped into a black box.
However, this forces Chisel to use Verilator instead of Treadle to run the simulation, impacting
startup time.

Chisel does not fully support VHDL. It can support VHDL using VCS, but there is no
open-source solution available for VHDL simulation. For companies with a lot of source code written in VHDL this is a concern, as they must be able to integrate their existing IP in a Scala/Chisel based design and verification workflow.
All major commercial simulation and synthesis tools support mixed-language designs, but no open-source tools exist that provide the same functionality.

To alleviate this issue, we use the open-source Yosys synthesis suite \cite{Yosys}.
Yosys is an open-source digital hardware synthesis suite for Verilog.
Yosys also has a variety of plugins, one of these being a plugin for using GHDL \cite{ghdl},
an open-source VHDL simulator. By using Yosys in conjunction with GHDL,
VHDL files are compiled to an RTL-based intermediate representation,
which is then written to a Verilog file using Yosys.
GHDL has full support for IEEE 1076 VHDL 1987, 1993, 2002, and a subset of 2008.
A working solution named VHDL2Verilog has been made for this, which has been
tested with certain simple VHDL designs \cite{vhdl2verilog}.

This paper is an extension of \cite{verify:chisel:2020}.

In the following sections, we will explore the different backgrounds on which ChiselVerify was based.

\section{Background and State-of-the-Art}
\label{sec:background}

VHDL and Verilog are the classic hardware description languages, first appeared in the 1980s.
SystemVerilog~\cite{SystemVerilog}, as an extension to Verilog, adds features from VHDL
for the hardware description and object-oriented features for verification.
Recent advances with SystemVerilog and Chisel \cite{chisel:dac2012, chisel:book} have
brought object-oriented programming into the digital design and verification process.

Chisel is a ``Hardware Construction Language'' embedded in Scala, to describe digital circuits~\cite{chisel:dac2012}.
Scala/Chisel brings object-oriented and functional programming into the world of digital design.
For hardware generation and testing, the full Scala language and Scala and Java libraries are available.
As Scala and Java's full power is available to the verification engineer,
the verification process is also made more efficient.

Chisel is a hardware construction language embedded in Scala.
Chisel allows the user to write hardware generators in Scala, an object-oriented and functional language.
For example, we read in the string based schedules for a network-on-chip~\cite{s4noc:nocarc2019}
and convert them with a few lines of Scala code into a hardware table to
drive the multiplexer of the router and the network interface.

Chisel is solely a hardware \emph{construction} language, and thus all valid Chisel code
maps to synthesizable hardware.
By separating the hardware construction and hardware verification languages,
it becomes impossible to write non-synthesizable hardware and in turn, speeds up the design process.

Chisel and Scala are executing on the Java virtual machine and therefore have a very good
interoperability with Java. Therefore, we can leverage a large pool of Java libraries for
hardware design and  verification.
Furthermore, the name space of packets in Scala/Java simplifies integration of
external components.
Open source hardware components in Chisel can be organized like software
libraries at Maven servers.

SystemVerilog adds object-oriented concepts for the non-synthesizable verification code.
The SystemVerilog direct programming interface~\cite{Doulos:SV:dpi} allows the programmer to call
C functions inside a SystemVerilog (UVM) testbench.
This enables co-simulation with a ``golden model'' written in C, and the
testbench verifying the device under test (DUT).
With ChiselTest we can co-simulate with Java and Scala models and use the Java Native Interface
to co-simulate with models in C.

The Universal Verification Method (UVM) is an open source collection of SystemVerilog,
which is becoming popular in industry.
SystemVerilog has become a complex language with more than 250 keywords, and it is unclear
which tools support which language constructs for hardware description.
In contrast with Chisel, when the program compiles, it is synthesizable hardware.
Chisel is a small language, where the cheat sheet fits on two pages.
The power of Chisel comes from the embedding in Scala.
Furthermore, as classic hardware description languages are niche products, not
many tools or libraries are available. 
With Chisel on Scala we have the choice of different integrated development environments (IDE),
testing infrastructure (e.g., ScalaTest), and many free libraries.

The Java JNI (Java Native Interface) allows for a similar functionality in Java programs,
allowing them to call C functions and use their functionality.
By using Scala, which is built on Java, it is our hope to use the JNI together with Scala's test frameworks.
The aim is to develop a framework for co-simulation with Scala/Chisel testers and a
C-based golden model. This should allow companies to keep their existing C models,
but move their simulation workflow into Scala/Chisel testers.

The digital design described in Chisel can be tested and verified with
ChiselTest~\cite{chisel:tester2}, a non-synthesizable testing framework for Chisel.
ChiselTest emphasizes usability and simplicity while providing ways to scale up complexity.
Fundamentally, ChiselTest is a Scala library that provides access into the simulator through
operations like poke (write value into circuit), peek (read value from circuit, into the test framework), and step (advance time).
As such, tests written in ChiselTest are just Scala programs, imperative code that runs one line after the next.
This structure uses the latest programming language developments that have been implemented into Scala
and provides a clean and concise interface, unlike approaches that attempt to reinvent the wheel like UVM.

Furthermore, ChiselTest tries to enable testing best practices from software engineering.
Its lightweight syntax encourages writing targeted unit tests by making small tests easy.
A clear and clean test code also enables the test-as-documentation pattern,
demonstrating a module's behavior from a temporal perspective.

\section{Constraint Random Verification}
The complexity of digital design is growing with the capacity of the silicon. A decade ago, the industry started to move away from ``direct''
testing towards functional coverage and formal methods. One of the pillars of functional verification is constraint programming.
Constraint programming (CP) is a programming paradigm that has been developed since the mid-1980s and emerged as a further development of logic
programming. Constraint-based programming allows constraints and their solution mechanisms to be integrated into a programming language.
With constraint programming, the user describes the problem in a declarative way, while the solution process takes a back seat from the user's
perspective. A subset of these problems is the so-called Constraint Satisfaction Problems (CSP), which are mathematical problems defined as a
set of objects such that their state must satisfy several constraints. CSP represents the entities of a problem as a finite homogeneous
collection of constraints.

\begin{lstlisting}[caption={Random object in SystemVerilog}, label={lst:randobjsysv}]
typedef enum {UNICAST=11,MULTICAST,BROADCAST} pkt_type;

class frame_t;
    rand pkt_type ptype;
    rand integer len;
    rand bit  [7:0] payload [];
    constraint common {
        payload.size() == len;
    }
    // Constraint the members
    constraint unicast {
        len <= 2;
        ptype == UNICAST;
    }
    // Constraint the members
    constraint multicast {
        len >= 3;
        len <= 4;
        ptype == MULTICAST;
    }
endclass
\end{lstlisting}
Listing \ref{lst:randobjsysv} shows a class named "frame\_t". It uses the "rand" keyword for variables "len", "ptype", and payload.
Therefore these are the variables that can be randomized. Then constraints to these variables are applied and declared by the ``common'' "unicast,"
and "multicast" constraint groups. Each class in SystemVerilog has an intrinsic method called "randomize()," which causes new values to be selected
for all the variables declared with the rand keyword. The selected value for each variable will respect the constraints applied to it. If there are
``rand'' variables that are unconstrained, a random value inside their domain will be assigned. Combining random classes using the inheritance OOP
paradigm allows the creation of general-purpose models that can be constrained to perform domain-specific functions. In the research process, a
CSP solver was implemented in Scala based on the method described in \cite{russell2002artificial}. The implementation is composed of two main components.
The first one is the CSP solver itself, which uses a combination of backtracking and arc consistency to generate solutions for well-defined problems.
The second component is a small DSL, which allows users to declare and randomize objects.

\begin{lstlisting}[language=scala, caption={Random object in Scala}, label={lst:randobjscala}]

object pktType extends SVEnumeration {
    val UNICAST: Value = Value(11)
    val MULTICAST: Value = Value(0)
    val BROADCAST: Value = Value(1)
    
    val domainValues = {
      values.map(x => x.id).toList
    }

}
class Frame extends Random {
  import pktType._
  var pType: RandInt = 
        rand(pType, pktType.domainValues())
  var len: RandInt = 
        rand(len, 0 to 10 toList)
  var noRepeat: RandCInt = 
        randc( noRepeat, 0 to 1 toList)
  var payload: RandInt =
        rand(payload, 0 to 7 toList)

  val common = constraintBlock (
    binary ((len, payload) => len == payload)
  )

  val unicast = constraintBlock(
    unary (len => len <= 2),
    unary (pType => pType == UNICAST.id)
  )

  val multicast = constraintBlock(
    unary (len => len >= 3),
    unary (len => len <= 4),
    unary (pType => pType == MULTICAST.id)
  )
}
\end{lstlisting}

Listing \ref{lst:randobjscala}, shows an example of random a object. Contrary to SystemVerilog, to declare a random object, the user has to extend
the class from the Random base-class provided by the library. After that each random variable has to be declared of type "RandInt" and initialized
with the "rand" macro. Finally, like for SystemVerilog, in Scala inheriting the Random base class exposes the method "random" which assigns random
values to the random fields of the class.

\section{Verification of AXI4 Interfaced Components}

Another solution to the ever-increasing complexity of digital designs is to use standardized interfaces,
which enable greater reuse. One of such standard interfaces is AXI4, an open standard by ARM \cite{axi4standard},
which is used in particular to connect processor nodes to memories.
As such, most available synthesis tools, including Xilinx' Vivado,
provide IP generators whose output IP blocks are equipped with AXI interfaces
along with optional verification structures written in (System-)Verilog \cite{axi4vip}.

Typically, verification of components with such standard interfaces is provided through so-called \textit{bus functional models} (BFMs) that abstract complex low-level signal transitions between bus masters and slaves to a transaction level (e.g., write and read transactions). Unfortunately, such BFMs are not yet available in Chisel -- hence, why we include an example BFM based around ChiselTest in our framework.

\subsection{Introduction to AXI4}
The Advanced eXtensible Interface protocol by ARM is a highly flexible interconnect standard based around five separate channels; three for write operations and two for read operations. Operations, known as transactions, consist of a number of transfers across either set of channels. All channels share a common clock and active-low reset and base their transfers on classic ready-valid handshaking. It is designed with DMA-based memories in focus supporting multiple outstanding transactions and out-of-order completion. The five channels are:
\begin{itemize}
  \item \textit{Write Address} for transferring transaction attributes from master to slave
  \item \textit{Write Data} for transferring write data and strobe from master to slave
  \item \textit{Write Response} for transferring transaction status of a writes from slave to master
  \item \textit{Read Address} same as \textit{Write Address}, but for reads
  \item \textit{Read Data} for transferring read data from slave to master
\end{itemize}

Consider for example a write transaction of 16 data elements. First, the master provides transaction attributes (e.g., target address, burst length, and data size) as a single transfer over the \textit{Write Address} channel, then the master transfers the 16 data elements one at a time over the \textit{Write Data} channel, and finally, the slave indicates the status of the transaction over the \textit{Write Response} channel. The \textit{Read Address} and \textit{Read Data} channels may operate independently at the same time. A full description is available in \cite{axi4standard}. 

\subsection{Implementation}
Our implementation includes bundles definining the five different channels, abstract classes representing both master and slave entities, transaction-related classes, and of course the BFM itself; the \texttt{FunctionalMaster} class. The BFM is parameterized with a DUT that extends the slave class and provides a simple, transaction level interface to control the DUT. As such, its two most important public methods are \texttt{createWriteTrx} and \texttt{createReadTrx} which do exactly as their names indicate; create and enqueue write and read transactions. \\

Internally, the BFM makes use of ChiselTest's multithreading features to allow for (a) non-blocking calls to the aforementioned methods (i.e., one can enqueue multiple transactions without waiting for their completion) and (b) emulating the channel independence more closely. As such, when, for example, a write transaction is enqueued and no other write transactions are in-flight, the BFM spawns three new threads, one for each required channel. The threads each handle the handshaking necessary to operate the channels.

\subsection{A Simple Example}
Returning to the example used before, using the BFM to test a module called \texttt{Memory} is as simple as shown below. Creating a write transaction with 16 data elements (minimum burst length is 1, hence \texttt{len = 15} means a burst of 16 items) takes just one call to a method the majority of whose arguments have default values. It is equally simple to create a subsequent read transaction -- but beware that due to the BFM's parallel execution style, the channels are indeed independent. As such, not waiting for a write to complete before starting to read from the same address may return incorrect results depending on the implementation of the DUT.
\begin{lstlisting}[language=scala, caption={Using the AXI4 BFM with ChiselTest}, label={lst:axitest}]
class MemoryTester extends FlatSpec with ChiselScalatestTester with Matchers {
  behavior of "My Memory module"
  it should "write and read" in {
    test(new Memory()) {
      dut =>
        val bfm = new FunctionalMaster(dut)
        master.createWriteTrx(0, Seq.fill(16)(0x7FFFFFFF), len = 15, size = 2)
        master.createReadTrx(0, len = 15, size = 2)
    }
  }
}
\end{lstlisting}

\section{Coverage in Chisel}

One of the main tools used in verification is test coverage. This allows verification engineers to measure their progress throughout the testing process and have an idea of how effective their tests actually are. Coverage can be separated into two distinct categories: code coverage and functional coverage. code coverage defines a quantitative measure of the testing progress, \textit{"How many lines of code have been tested?"}, whereas functional coverage gives a rather qualitative measure, \textit{"How many functionalities have we tested?"}.  Our solution gives the verification engineer access to two ways of obtaining their code coverage and new constructs allowing the definition of a verification plan and the creation of a functional coverage report directly integrated into the Chisel testing framework.

\paragraph{Code Coverage with Treadle}  
The first part of our solution is about code coverage, more specifically line coverage that was added to the Treadle FIRRTL execution engine. Treadle is a common FIRRTL execution engine used to simulate designs implemented in Chisel. This engine runs on the FIRRTL intermediate representation code generated by a given Chisel implementation and allows one to run user-defined tests on the design using frameworks like \textit{iotesters} or the more recent \textit{testers2}. In our pursuit of creating a verification framework, we found that one way to obtain line coverage would be to have our framework run on an extended version of Treadle that was capable of keeping track of said information.

The solution that was used to implement line coverage was based off of a method presented by Ira. D. Baxter~\cite{branch-cov-made-easy:2002}. The idea is to add additional outputs for each multiplexer in the design. These new ports, which we will call \textit{Coverage Validators}, are set depending on the paths taken by each multiplexer and that information is then gathered at the end of each test and maintained throughout a test suite. Once the testing is done, we used the outputs gathered from the \textit{Coverage Validators} to check wether or not a certain multiplexer path was taken during the test, all of this resulting in a branch coverage percentage.

This was implemented in Treadle by creating a custom pass of the FIRRTL compiler that traverses the Abstract Syntax Tree (\textit{AST}) and adds the wanted outputs and coverage expressions into the source tree. Once that is done, the \texttt{TreadleTester} samples those additional outputs every time the \texttt{expect} method is called and keeps track of their values throughout a test suite. Finally it generates a Scala \texttt{case class} containing the following coverage information:
\begin{itemize}
\item The multiplexer path coverage percentage.
\item The coverage Validator lines that were covered by a test.
\item The modified LoFIRRTL source code in the form of a \texttt{List[String]}.
\end{itemize}
The \texttt{CoverageReport} case class can then be serialized, giving the following report:
\begin{verbatim}
COVERAGE: 50.0% of multiplexer paths tested
COVERAGE REPORT:

+ circuit Test_1 :
+   module Test_1 :
+     input io_a : UInt<1>
+     input io_b_0 : UInt<2>
+     input io_b_1 : UInt<2>
+     input clock : Clock
+     output io_cov_valid_0 : UInt<1>
+     output io_cov_valid_1 : UInt<1>
+     output out : UInt<2>
+   
+     io_cov_valid_0 <= io_a
-     io_cov_valid_1 <= not(io_a)
+     out <= mux(io_a, io_b_0, io_b_1)
\end{verbatim}
The example above is taken for a simple test, where we are only testing the path where \texttt{in\_a} is 1. This means that, since we only have a single multiplexer, only half of our branches have been tested and we would thus want to add a test for the case where \texttt{in\_a} is 0. The report can thus be interpreted as follows:  
\begin{itemize}
\item "\texttt{+}" before a line, means that it was executed in at least one of the tests in the test suite.
\item "\texttt{-}" before a line, means that it wasn't executed in any of the tests in the test suite.\\
\end{itemize}

Treadle thus allows us to obtain coverage at the FIRRTL level. A more interesting result would be if the FIRRTL line coverage would be mapped to the original Chisel source. This is possible but challenging, since Treadle only has access to the original source code through \textit{Source locators} which map some of the FIRRTL lines back to Chisel. This means that the code can only be partially mapped and the remainder will have to be reconstructed using some smart guessing.\\

\paragraph{Functional Coverage Directly in Scala}
Functional Coverage is on the principal tools used during the verification process, since it allows one to have a measurement of \textit{"how much of the specification has been implemented correctly"}. A verification framework would thus not be complete without constructs allowing one to define a verification plan and retrieve a functional coverage report. The main language used for functional coverage is \textit{SystemVerilog}, which is why our solution is based on the same syntax. There are three main components to defining a verification plan: 
\begin{itemize}
\item \texttt{Bin}: Defines a range of values that should be tested for (i.e. what values can we expect to get from a given port).
\item \texttt{CoverPoint}: Defines a port that needs to be sampled in the coverage report. These are defined using a set of bins.
\item \texttt{CoverGroup}: Defines a set of \texttt{CoverPoint}s that need to be sampled at the same time.
\end{itemize}
Using the above elements, one can define what's known as a verification plan, which tells the coverage reporter what ports need to be sampled in order to generate a report.
In order to implement said elements in Scala we needed to be able to do the following:
\begin{itemize}
\item Define a verification plan (using constructs similar to \texttt{coverpoint} and \texttt{bins}).
\item Sample DUT ports (for example by hooking into the \textit{Chisel Testers2} framework).
\item Keep track of bins to sampled value matches (using a sort of DataBase).
\item Compile all of the results into a comprehensible Coverage Report.
\end{itemize}
Implementing these elements was done using a structure where we had a top-level element, known as our \texttt{Coverage Reporter} which allows the verification engineer to define a verification plan using the \texttt{register} method, which itself stores the \texttt{coverpoint} to \texttt{bin} mappings inside of our \texttt{CoverageDB}. Once the verification plan is defined, we can sample our ports using the \texttt{sample} method, which is done by hooking into \textit{Chisel Testers2} in order to use its peeking capabilities. At the end of the test suite a functional coverage report can be generated using the \texttt{printReport} method, which shows us how many of the possible values, defined by our bin ranges, were obtained during the simulation.
\begin{lstlisting}[language=scala]
val cr = new CoverageReporter
cr.register(
    //Declare CoverPoints
    //CoverPoint 1
    CoverPoint(dut.io.accu , "accu",
        Bins("lo10", 0 to 10)::
        Bins("First100", 0 to 100)
        ::Nil)::
    //CoverPoint 2
    CoverPoint(dut.io.test, "test", 
        Bins("testLo10", 0 to 10)
        ::Nil)::
    Nil,
    //Declare cross points
    Cross("accuAndTest", "accu", "test",
        CrossBin("both1", 1 to 1, 1 to 1)
        ::Nil)::
    Nil)
\end{lstlisting}
The above code snippet is an example of how to define a verification plan using our coverage framework. The concepts are directly taken from \texttt{SystemVerilog}, so it should be accessible to anyone coming from there. One concept, that is used in the example verification plan, which we haven't presented yet is the idea of \textit{Cross Coverage} defined using the \texttt{Cross} construct. \textit{Cross Coverage} allows one to specify coverage relations between CoverPoints. This means that a cross defined between, let's say, \texttt{coverpoint a} and \texttt{coverpoint b} will be used to gather information about when \texttt{a} and \texttt{b} had certain values simultaneously. Thus in example verification plan we are checking that \texttt{accu} and \texttt{test} take the value 1 at the same time.\\
Once our verification plan is defined, we need to decide when we want to sample our cover points. This means that at some point in our test, we have to tell our \texttt{CoverageReporter} to sample the values of all of the points defined in our verification plan. This can be done, in our example, simply by calling \texttt{cr.sample()} when we are ready to sample our points. Finally once our tests are done, we can ask for a coverage report by calling \texttt{cr.printReport()} which results in the following coverage report: 
\begin{verbatim}
============== COVERAGE REPORT ==============
================ GROUP ID: 1 ================
COVER_POINT PORT NAME: accu
BIN lo10 COVERING Range 0 to 10 HAS 8 HIT(S)
BIN First100 COVERING Range 0 to 100 HAS 9 HIT(S)
============================================
COVER_POINT PORT NAME: test
BIN testLo10 COVERING Range 0 to 10 HAS 8 HIT(S)
============================================
CROSS_POINT accuAndTest FOR POINTS accu AND test
BIN both1 COVERING Range 1 to 1 CROSS Range 1 to 1 
HAS 1 HIT(S)
============================================
\end{verbatim}
An other option would be, for example if we want to do automated constraint modifications depending on the current coverage, to generate the coverage as a Scala \texttt{case class} and then to use it's \texttt{binNcases} method to get numerical and reusable coverage results.  
  
One final element that our framework offers is the possibility to gater delayed coverage relationships between two coverage points. The idea is similar to how a \texttt{Cross} works, but this time rather than sampling both points in the same cycle, we rather look at the relation between one point at the starting cycle and an other point sampled a given number of cycles later. This number of cycles is called the \texttt{delay} and there are currently three different ways to specify it:  
\begin{itemize}
 \item \texttt{Exactly} delay, means that a hit will only be considered if the second point is sampled in its range a given number of cycles after the first point was.
 \item \texttt{Eventually} delay, means that a hit will be considered if the second point is sampled in its range at any point within the following given number of cycles after the first point was.  
 \item \texttt{Always} delay, means that a hit will be considered if the second point is sampled in its range during every cycle for a given number of cycles after the first point was sampled.
\end{itemize}

\section{Use Case: Hardware sorting}

In the process of the research a use case provided by Microchip was implemented in order to apply developed testing and verification features.
In the following section the implementation of the use case and the connected testing will be discussed. The code can be found in the 
\href{https://github.com/chisel-uvm/chisel-verify/tree/master/src/main/scala/heappriorityqueue}{project repository}.

\subsection{Specification}

The provided specification document describes a priority queue which can be used in real time systems for the scheduling of deadlines by providing information about the 
next timer expiration to the host system. Sorting of the enqueued elements is conducted by applying the heap sort algorithm. Elements are structured in a so-called 
heap which is a tree data structure. The tree needs to be balanced in order for the
timer closest to expiring to get to the top of the tree and thus to the head of the queue. This means verifying that every parent node is smaller than the connected child nodes.

The $\log_k N$ deepness of the tree provides good scalability in terms of insertion and removal times when the queue size increases, 
since worst case $\log_k N-1$ swap operations need to be conducted in order to rebalance the heap. Here $k$ is the number of child elements per parent and $N$
is the number of elements in the heap. A trade-off is the varying delay connected to the rebalancing of the tree where the queue is unresponsive. If queuing
happens in bursts, a buffer could be added. Here the introduced delay from insertion request to actual appearance of the enqueued value in the heap of course needs
to be taken into account.

In order for the host system to have the ability to distinguish between multiple consecutive super cycles and clock cycles in a super cycle, the values inserted 
into the queue are split into the fields \textit{cyclic} and \textit{normal} priority (time out value). The removal functionality of the queue requires a reference system. A reference 
ID is therefore given together with the element at insertion, where ID generation and uniqueness are handled by the host system.

\subsection{Implementation}

The implemented priority queue is described in chisel.
It is split into 3 modules: The \texttt{Heapifier}, responsible for the sorting, the \texttt{QueueControl}, taking care
of the general control flow in the queue and the \texttt{Memory} module which handles memory accesses and can search the memory for a specific reference 
ID.

In order for the priority queue to work efficiently it is crucial to optimize memory accesses. Therefore a layout is proposed in the specification where all 
child elements of a certain node are stored together under one memory address. This allows single memory access fetches of all $k$ children. Since 
the root node has no siblings it is stored alone in a register. This enables even faster access in certain scenarios which are discussed later on.

One memory row contains $k$ elements each consisting of the 3 introduced fields: \textit{cyclic} priority, \textit{normal} priority and the connected
reference ID. In the implemented 
\texttt{Memory} module a single sequential memory is instantiated where masking is used to be able to over-write specific elements in one memory row.

There are a variety of solutions to the problem of content addressability which is required here in order to find positions of elements in the heap by providing
their reference ID. Cache-like memory relying on parallel tag comparison could be used to achieve fast and constant search times. On the other hand, the 
existing sequential memory could be searched linearly, where $k$ reference ID's are compared in parallel until the searched reference ID is found.
A compromise between the two solutions could include splitting memory space over multiple instances of the latter and thus reducing worst case search time.
The priority queue is designed to be independent of the specific implementation. As a reference, the linear search is implemented.

\begin{figure}
	\centering
	\includegraphics[width=0.35\textwidth]{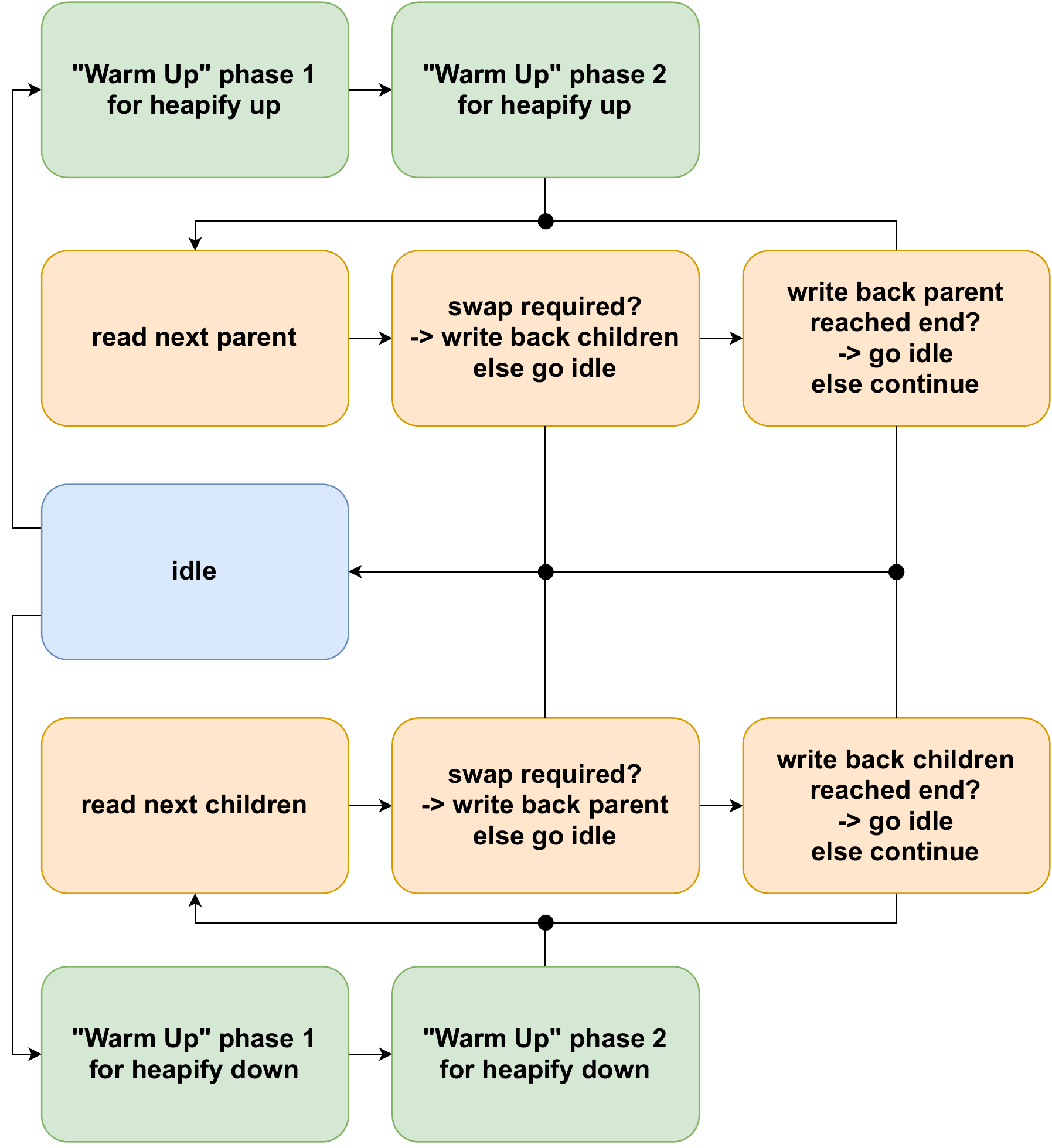}
	\caption{The state diagram of the \texttt{Heapifier}}
	\label{fig:pq_heapifier_state}
\end{figure}

The \texttt{Heapifier} loops from a given starting point in the tree either upwards or downwards until it hits the root or a childless node. In each iteration 
it is checked whether the parent element is smaller than its child elements and if not a swap occurs. Once the parent element and child elements of the starting 
point are fetched from memory, only the next parent or block of children respectively needs to be fetched depending on the looping direction (up/down). Thus 
only 3 cycles (1 fetch, 2 write backs) are required per swap. The state diagram of the \texttt{Heapifier} is shown in Figure \ref{fig:pq_heapifier_state}.

\begin{figure}
	\centering
	\includegraphics[width=0.45\textwidth]{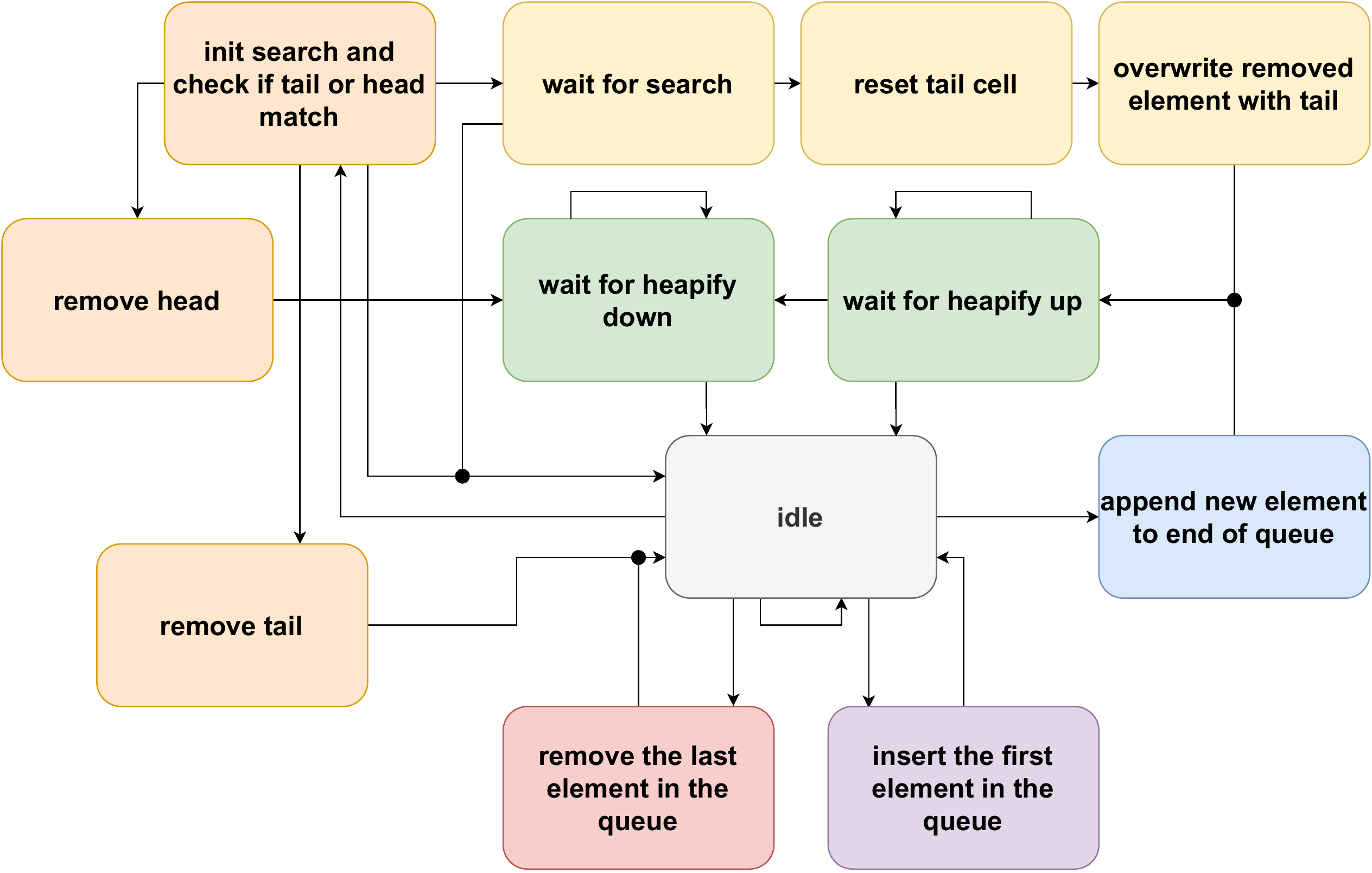}
	\caption{The state diagram of the \texttt{QueueControl}}
	\label{fig:pq_control_state}
\end{figure}

The task of the \texttt{QueueControl} is to insert or remove elements and then signalize to the \texttt{Heapfier} to balance the tree. As it can be seen in Figure 
\ref{fig:pq_control_state}, there are a series of special cases where insertion or removal times can be reduced for instance by taking advantage of the head element
being saved in a register. The achieved best and worst case insertion as well as removal times are presented in Table \ref{tab:pq_timings}.

\begin{table}
	\centering
	\begin{tabular}{l|c}
		Head insertion & 2 cycles \\
		Normal insertion & min 7 cycles and max $5+3\cdot \mathrm{log}_k(N)$\\
		Head removal & min 8 cycles and max $6+3\cdot \mathrm{log}_k(N)$\\
		Tail removal & 3 cycles\\
    Normal removal & \begin{tabular}{c}min 12 cycles + search time\\ max $13+3\cdot \mathrm{log}_k(N)$ + search time \end{tabular} \\
    \multicolumn{2}{c}{\scriptsize{($N=$ queued elements, $k=$ number of child elements per node)}}
	\end{tabular}
	\caption{Best and worst case insertion and removal times}
	\label{tab:pq_timings}
\end{table}
\begin{table}[]
  \centering
  \begin{tabular}{l|cccc}
      Size &$k=2$&$k=4$&$k=8$&$k=16$\\\hline
      16+1 & 9.38/14.56&7.86/11.51&7.47/10.42&-\\
      32+1 & 9.9/15.15&7.96/12.29&7.34/11.22&7.54/11.00\\
      64+1 & 9.79/16.34&8.01/13.80&7.41/12.38&7.32/11.18\\
      128+1 & 9.96/17.37&8.14/14.47&7.54/13.14&7.26/11.62\\
      256+1 & 9.73/17.39&8.15/15.54&7.53/14.21&7.34/12.89
  \end{tabular}
  \label{tab:pq_simres}
  \caption{Simulated insertion/removal times.}
\end{table}

\subsection{Testing and Verification}

All modules and the fully assembled queue were tested with random stimuli in scala by employing the ChiselTest framework.
In order to check whether the dut matched the specification, reference models were written for each module. Most modules 
could be modelled by a single or multiple functions. As a reference model for the whole priority queue, a class 
was written which simulates state and interaction on a operation based level. In order to abstract interaction with the DUT,
wrapper classes were employed. These make it easy to think on a transaction or 
operation based level when writing tests.

In the test of the priority queue, purely random pokes are produced. In order to evaluate how well these pokes were spread over the spectrum 
of possible or interesting input combinations, the developed functional coverage feature is used. This allows to evaluate whether interesting 
or important edge cases are reached by the random input sequence. Furthermore, metrics on how many of the generated pokes actually are valid 
operations, are collected. The average insertion and removal times measured under a random test run are shown in table \ref{tab:pq_simres}.
These numbers are of course heavily dependent on access patterns and are as such only representative for the completely random test case used here.
The tests can be found in the \href{https://github.com/chisel-uvm/chisel-verify/tree/master/src/test/scala/heappriorityqueue}{project repository}.

\section{UVM, SystemVerilog and Verification}
This section serves as a reference on what the UVM can do, how it works, and how some of this (namely the SystemVerilog Direct Programming Interface (DPI)) may be implemented in a Scala based testing framework.

The Universal Verification Methodology (UVM) is a verification framework built on top of the testing capabilities of SystemVerilog. Prior to the introduction of the UVM, the major EDA simulation vendors supported different frameworks / methodologies. This meant that a verification environment designed for one simulator might not run under another simulator. The main purpose of the UVM was to standardize the testing framework used across EDA vendors, making it easier for users to use their testbenches across different software suites.

The testbench in the UVM is built up around a number of components.
Each component performs only one task in the testbench, allowing the engineer to make changes to some components without affecting others. 
For example, the sequencer is the component responsible for generating transactions for the DUT, whereas the driver is responsible for converting the transaction into pin-level wiggles, i.e., generating correct start/stop conditions and
driving signals. If a new sequence of transactions is to be generated, only the sequencer is affected. Likewise, the sequencer does not care how the transactions are converted into pin-level signals---this is the sole responsibility of the driver. This separation into several components results in a more structured testbench design as there are fewer dependencies than in a monolithic testbench.

The main components of a UVM testbench are as follows:

    A \textit{Sequence(r):} defines the order of transactions necessary for a given purpose, e.g., synchronization or reset sequence. The sequencer is responsible for transferring the transactions, defined by the sequence, to the driver.
    
    A \textit{Driver} converts transactions into pin-level signals and drives these signals onto the DUT.
    
    An \textit{Interface} is a SystemVerilog construct which allows the user to group related signals. A DUT may have several interfaces attached.
    The interface is used to avoid hooking directly into the DUT, making it easier to test multiple DUT versions.
    
    A \textit{Monitor} monitors all traffic on the interface, converting pin-level signals into transaction-level objects that can be operated on by other components, such as a coverage collector or scoreboard.
    
    An \textit{Agent} encapsulates monitor, sequencer and driver, setting configuration values. Agents may be set active or passive (with or without a driver and sequencer). An agent is useful when it is necessary to have multiple instances of the same components, e.g., when a 4-port network switch needs four identical drivers with different configurations.
    
    A \textit{Scoreboard} is used to check whether correct functionality is achieved. Usually does so by using a ``golden model'' for co-simulation via the SystemVerilog direct programming Interface.
    
    The \textit{Environment} is used to configure and instantiate all child components. Environments are typically application-specific and may be modified by the test.
    
    The \textit{Test} is the top-level verification component. The test designer may choose to perform factory overrides of classes and set configuration values here, which modify the child components.

As shown above, even a ``Hello, World'' example using the UVM requires that the user understands how and why each of the different UVM components should be used. The use of so many components causes UVM to have a very steep learning curve, which may discourage adoption. This also means that UVM is not the proper testing methodology for small designs or one-off tests due to the initial workload.
However, once the initial setup of the testbench is finished for large and complex designs, generating new tests becomes easier.

\subsection{Scoreboard and DPI}
The purpose of the scoreboard is to ensure that the DUT matches specification. When using directed tests (i.e. hand-written tests meant to test a single part of the specification), this may be as simple as constructing a list of input/output values and checking these in order. When using randomized testing, the scoreboard is usually implemented as a software model (sometimes called a golden model) which is defined to be correct. The software model should exactly mimic the functionality of the DUT, and thus provide the same outputs given the same inputs. 

The scoreboard may be implemented purely in SystemVerilog, or it may be implemented in a C-like language (C, C++ or SystemC). One of the benefits that \SV adds to the verification environment is the ability to interface with C-code through the use of the \SV Direct Programming Interface (DPI). This is an interface which allows C code to be called from inside the \SV testbench, and likewise allows for \SV code to be called from inside of a C program. Programming low-level simulations of hardware is most likely easier in C than in plain \SV. In listing \ref{lst:dpi} is shown a simple example of some SystemVerilog code calling C code through the \SV DPI. 

Notice the inclusion of the header file \texttt{svdpi.h}, which contains the definitions necessary for interoperability. Once this is done, the function must be imported in \SV by use of the \texttt{import "DPI-C"} statement, after which the function may be called as any other \SV function. Using the DPI is surprisingly simple and painless, and makes it very simple to integrate a C model in the testbench.

It should be noted that a scoreboard doesn't necessarily "rate"\, the performance of the DUT by comparing it to other modules, as one might expect from the name. The DUT is only compared against the reference model, and the "rating" is how many assertions pass/fail in a given test.

\longlist{snippets/hello.c}{Short example showing how to use the \SV DPI to execute C-code from within \SV.}{lst:dpi}

\subsection{Constrained Random Verification}

Most UVM testbenches employ "Constrained Random Verification"\, (CRV) for generating test stimulus. This is as opposed to using directed testing, where input vectors are predefined to test a certain behaviour. With CRV, random test stimuli are generated. The "Constrained" part of CRV implies that these values aren't entirely random, but are chosen to fulfill a series of criteria. If eg. the DUT is an ALU with a 5-bit opcode field of which only 20 bit-patterns are used, it would be prudent to only generate the 20 valid opcodes for most test purposes. Specific tests could then be written to ensure proper operation if an invalid opcode was asserted.

An example of a class with randomized variables is seen in listing \ref{lst:crv}. 

The keyword \texttt{constraint} is used to constrain a variable defined with the \texttt{rand} keyword.
One randomized variable, \texttt{bcd}, is a 4-bit field which only takes on values in the range 0 and 9. 
The field \texttt{value} has a $1/3$ chance of being 0, $1/3$ of being 255 and $1/3$ of being any other value. 
The fields \texttt{a,b,c} must satisfy $0<a<b<c$. 
The field \texttt{op} has no constraints, and will thus randomly toggle between  0, 1, 2 and 3, the only values it can take on.

\longlist{snippets/Myclass.svh}{Example SystemVerilog code showing how different values are constrained.}{lst:crv}

In listing \ref{lst:usingcrv}, an object of type \texttt{Myclass} is instantiated and randomized using the \texttt{mc.randomize()} command. Finally, the \SV function \texttt{\$display} is used to print the value of the BCD field. The \texttt{randomize} keyword is a \SV construct which will try to randomize all random fields of given class.


\longlist{snippets/top.sv}{Example SystemVerilog code showing how to instantiate and randomize a class with random fields.}{lst:usingcrv}

\subsection{Coverage collection}
The purpose of coverage collection, in this case functional coverage collection, is to check whether all "necessary"\, (as defined by the verification engineer) input combinations have been generated. For the ALU mentioned above, it might be interesting to check whether all opcodes work correctly when asserted after a reset, and whether over/underflow flags are correctly set when performing arithmetic operations. In general, functional coverage is concerned with evaluating whether all functionality of a device has been tested. This is opposed to line coverage, which evaluates which lines of code were run during a simulation.

\longlist{snippets/Cover.svh}{Examle SystemVerilog code showing how covergrups and coverpoints are organized.}{lst:cov}

An example of functional coverage collection is seen in listing \ref{lst:cov} where the randomized values from before are covered. A \textit{covergroup} is a label used to collect relevant values under the same heading. A \textit{coverpoint} is a directive instructing the simulator that the given value should be monitored for changes. In the declaration of the coverpoint, several \textit{bins} are generated. These bins correspond to the bins of a histogram. Every time an event occurs which matches the declaration inside a bin, the counter associated with that bin is incremented by one. An event may cause multiple bins to increment at the same time.

In the coverage declarations shown in listing \ref{lst:cov}, the covergroup \texttt{cg\_bcd} only covers one field, \texttt{bcd}. The bin is labeled by prepepending \texttt{BCD:} in front of the \textit{coverpoint} directive. 10 bins are generated which each sample the values 0-9, and the remaining values 10-15 are sampled into the default bin \texttt{others}. 

In the covergroup \texttt{cg\_others} three coverpoints are set up. The \texttt{VAL} coverpoint samples 3 ranges of values. Any value in the range \texttt{[0:20]}will cause the counter on bin \texttt{low} to increment by one. Likewise for the other bins in that coverpoint. The \texttt{A} coverpoint auto-generates one bin for each possible value it may take on, 16 bins in total, since no bins are explicitly declared. The coverpoint \texttt{OP} has one bin, \texttt{toggle} which only increments when \texttt{mode} toggles from \texttt{0x0} to \texttt{0x1}. Finally, the \texttt{cross} statement implements cross coverage. Cross coverage tracks what values were sampled at multiple coverpoints at the same time.

Using the cross of \texttt{A} and \texttt{OP}, it may be possible to have 100\% coverage on both coverpoints (ie. all bins have been hit at least once), but the cross coverage may not be at 100\% if eg. \texttt{OP} never toggled while \texttt{A} was 1. Increasing the number of random samples that are generated may alleviate this problem. If it doesn't, it may be indicative that something is wrong in the structure of the testbench or DUT.

In listing \ref{lst:usingcov}, the module from \cref{lst:usingcrv} has been expanded to also use the coverage collector.

\longlist{snippets/top.sv}{Showcasing how multiple random values are generated and sampled by the coverage collector.}{lst:usingcov}

\begin{figure}[htbp]
	\centering
	\includegraphics[width=\columnwidth]{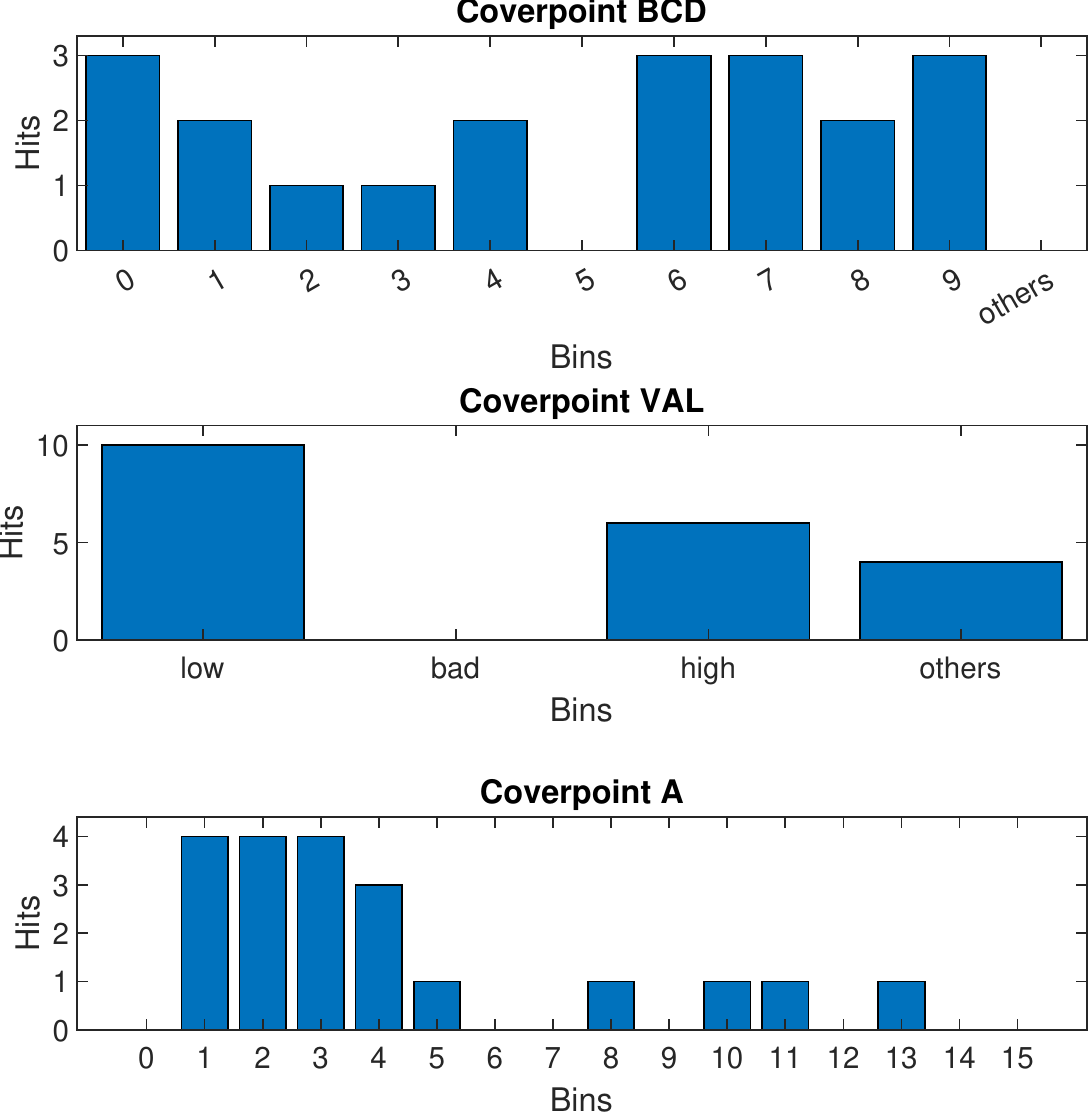}
	\caption{Coverage bins generated by running the code in listing \ref{lst:usingcov}}
\label{fig:coverage}
\end{figure}

In Figure \ref{fig:coverage}, the result of running the 20 iterations is seen for coverpoints \texttt{BCD, A, VAL}. 

\subsection{C Integration in Scala}
The \SV DPI is a great addition to the verification workflow, as it allows the designer to easily implement a C model into the \SV testbench.
Similar functionality is available in Scala by leveraging the Java Native Interface (JNI). This is an interface which allows Java / Scala programs to call native code, i.e. code compiled for the specific CPU it is running on. This is typically encountered as .DLL files on Windows or .so files on linux.

When using Chisel and other external libraries, it is recommended to use the Scala Build Tool (sbt) to manage the build. However, doing this makes it more difficult to use the default JNI integration in Scala. This can be alleviated by use of the plugin \texttt{sbt-jni} by Jakob Odersky. 

Listing \ref{lst:jni-scala} shows an example of a Scala file with a native function.

\longlist{snippets/Myclass.scala}{Example Scala code showing how to integrate native code in Scala.}{lst:jni-scala}

The annotation \texttt{@native} informs the Scala compiler that the function should be found in native code. The annotation \texttt{@nativeLoader} is necessary for use with the plugin. The name "native0"\, is the name of the current SBT project, appended with a 0.

Once the above file has been written, the corresponding C header file is generated by running \texttt{sbt javah}. The contents of the generated header file should then be copied into a C code file, where the functions can be implemented. In this case, it may look like the contents of listing \ref{lst:jni-c}.

\longlist{snippets/Myclass.c}{C implementation of the methods declared in Myclass from listing \ref{lst:jni-scala}.}{lst:jni-c}

Notice that primitives such as \texttt{int} can be autocast from a \texttt{jint}, whereas a string must be obtained used using the function pointers defined in the \texttt{JNIEnv} structure. Once this is done, a CMake makefile is generated by running \texttt{sbt "nativeInit cmake"}, and the C files are compiled using \texttt{sbt nativeCompile}. If all goes well, the code can then be run with \texttt{sbt run}. 

If changes are made to any of the function definitions or Scala file, no other steps are necessary than running \texttt{sbt run} again. This will also invoke the CMake script generation and compilation steps if necessary. If new native methods are added to the Scala file, \texttt{sbt javah} must be run again to generate new C headers.

For more information regarding the plugin setup, see the \href{https://github.com/jodersky/sbt-jni}{plugin page on Github} or the file HowToJni.md in the \href{https://github.com/chisel-uvm/documentation}{documentation repository}.

\section{Conclusion}

In this paper, we introduced ChiselVerify, an open-source solution that should increase a verification engineer's productivity by following the trend of moving towards a more high-level and software like ecosystem for hardware design. With this, we brought functional coverage, statement coverage, constraint random verification and transactional modelling to the Chisel/Scala ecosystem, thus allowing for the improvement of current engineer's efficiency and easing the way for software engineers to join the hardware verification world.

\subsection*{Acknowledgment}

This work has been performed as part of the
``InfinIT -- Innovationsnetv{\ae}rk for IT'', UFM case no. 1363-00036B,
``High-Level Design and Verification of Digital Systems''.

\bibliographystyle{abbrv}
\bibliography{msbib,chisel-uvm,testing}

\end{document}